\documentclass{sigchi}

\toappear{\scriptsize Permission to make digital or hard copies of all or part of this work for personal or classroom use is granted without fee provided that copies are not made or distributed for profit or commercial advantage and that copies bear this notice and the full citation on the first page. Copyrights for components of this work owned by others than the author(s) must be honored. Abstracting with credit is permitted. To copy otherwise, or republish, to post on servers or to redistribute to lists, requires prior specific permission and/or a fee. Request permissions from permissions@acm.org. \\
{\emph{CHI '20, April 25--30, 2020, Honolulu, HI, USA.} } \\
\copyright~2020 Copyright is held by the owner/author(s). Publication rights licensed to ACM. \\
ACM ISBN 978-1-4503-6708-0/20/04\ ...\$15.00.\\
http://dx.doi.org/10.1145/3313831.3376398 }

\CopyrightYear{2020}
\setcopyright{acmlicensed}
\doi{https://doi.org/10.1145/3313831.3376398}
\isbn{978-1-4503-6708-0/20/04}
\conferenceinfo{CHI'20,}{April  25--30, 2020, Honolulu, HI, USA}
\acmPrice{\$15.00}

\usepackage{balance}       
\usepackage{graphics}      
\usepackage[T1]{fontenc}   
\usepackage{txfonts}
\usepackage{mathptmx}
\usepackage[pdflang={en-US},pdftex]{hyperref}
\usepackage{color}
\usepackage{booktabs}
\usepackage{textcomp}
\usepackage{xspace}

\usepackage{microtype}        
\usepackage{ccicons}          

\usepackage[nocompress]{cite} 
\usepackage{makecell}
\usepackage[labelformat=simple]{subcaption} 

\makeatletter
\newcommand\footnoteref[1]{\protected@xdef\@thefnmark{\ref{#1}}\@footnotemark}
\makeatother

\newcommand{\eg}{\textit{e.g.}\@\xspace}
\newcommand{\ie}{\textit{i.e.}\@\xspace}
\newcommand{\etal}{\textit{et al.}\@\xspace}

\def\plaintitle{Pedestrian Detection with Wearable Cameras for the Blind: A Two-way Perspective}
\def\plainauthor{Kyungjun Lee, Daisuke Sato, Saki Asakawa, Hernisa Kacorri, Chieko Asakawa}

\def\plainkeywords{wearable camera, accessibility, social acceptance, pedestrian detection, face recognition, crowdsourcing}

\makeatletter
\def\url@leostyle{%
  \@ifundefined{selectfont}{
    \def\UrlFont{\sf}
  }{
    \def\UrlFont{\small\bf\ttfamily}
  }}
\makeatother
\def\pprw{8.5in}
\def\pprh{11in}

\definecolor{linkColor}{RGB}{6,125,233}
\hypersetup{%
  pdftitle={\plaintitle},
 pdfauthor={\plainauthor},
  pdfkeywords={\plainkeywords},
  pdfdisplaydoctitle=true, 
  bookmarksnumbered,
  pdfstartview={FitH},
  colorlinks,
  citecolor=black,
  filecolor=black,
  linkcolor=black,
  urlcolor=linkColor,
  breaklinks=true,
  hypertexnames=false
}

\begin{document}

\title{\plaintitle}


\numberofauthors{1}
\author{%
    \alignauthor{Kyungjun Lee\textsuperscript{1}, Daisuke Sato\textsuperscript{2}, Saki Asakawa\textsuperscript{3}, Hernisa Kacorri\textsuperscript{1}, Chieko Asakawa\textsuperscript{2, 4}\\
        \affaddr{\textsuperscript{1}University of Maryland,
        \textsuperscript{2}Carnegie Mellon University,
        \textsuperscript{3}New York University,
        \textsuperscript{4}IBM Research}
        \email{kjlee@cs.umd.edu, daisukes@cmu.edu, saki.asakawa@nyu.edu, hernisa@umd.edu, chiekoa@cs.cmu.edu}}\\
}




\maketitle
\begin{abstract}
Blind people have limited access to information about their surroundings, which is important for ensuring one's safety, managing social interactions, and identifying approaching pedestrians. With advances in computer vision, wearable cameras can provide equitable access to such information. However, the always-on nature of these assistive technologies poses privacy concerns for parties that may get recorded.  We explore this tension from both perspectives, those of sighted passersby and blind users, taking into account camera visibility, in-person versus remote experience, and extracted visual information.  We conduct two studies: an online survey with MTurkers (N=206) and an in-person experience study between pairs of blind (N=10) and sighted (N=40) participants, where blind participants wear a working prototype for pedestrian detection and pass by sighted participants.  Our results suggest that both of the perspectives of users and bystanders and the several factors mentioned above need to be carefully considered to mitigate potential social tensions.
\end{abstract}




\section{Introduction}
With advances in machine learning and computing devices that are small and light enough to be worn on one's body, wearable cameras hold the promise to increase real-world accessibility for people with visual impairments (\eg~\cite{zhao2019designing, al2016ebsar, sarfraz2017multimodal, stearns2018automated}) with few ideas making it to commercial products such as OrCam~\cite{OrCam2019}, Aira~\cite{ Aira2019} and eSight~\cite{eSight2019}. Given that these devices can capture information in the surroundings that people are uncomfortable sharing or that could be used for surveillance, public use of wearable cameras is conditional on societal consent. Thus, beyond technical challenges, such as battery life, accuracy, form factor, and split-attention, issues such as social acceptability and privacy concerns~\cite{profita2016effect} hold back
their use for more independent living.


\begin{figure}[!t]
    \centering
    \begin{subfigure}[t]{0.43\textwidth}
        \centering
        \includegraphics[width=\textwidth]{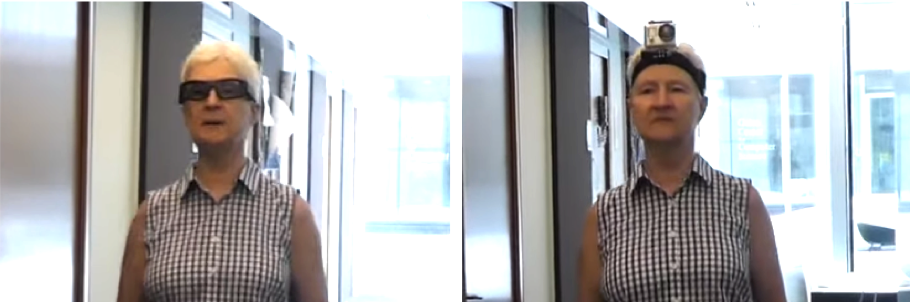}
        \caption{Passerby perspective video: A recording from the perspective of pedestrians watching a blind person who wears either smart glasses or a GoPro camera with a head strap, approaches, and walks by them.}
        \label{fig:video1}
    \end{subfigure}%
    \par\medskip
    \begin{subfigure}[t]{0.43\textwidth}
        \centering
        \includegraphics[width=\textwidth]{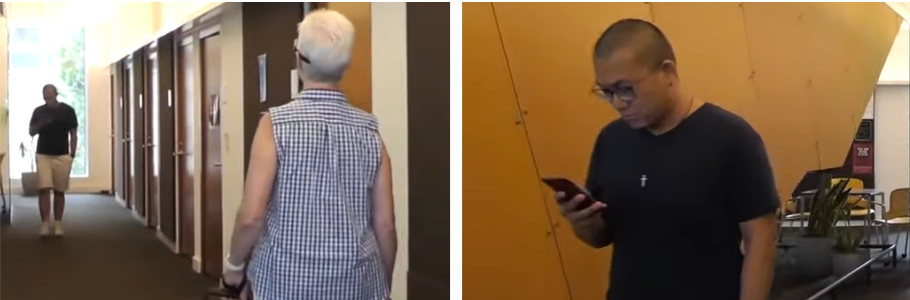}
        \caption{User perspective video: A recording showing a blind and a sighted person walking toward each other in a corridor. When in proximity, the perspective shifts to the wearable camera showing what it sees and what the blind user hears about the passerby behind the scenes.}
        \label{fig:video2}
    \end{subfigure}%
    \vspace{-0.07in}
    \caption{Video stimuli shown to our sighted participants during the online and in-situ studies. Both perspectives were included in the online study; in-situ only the user perspective.}
    \label{fig:video_desc}
    \vspace{-0.1in}
\end{figure}

With this paper, we aim to contribute to the ongoing discussion on the social tensions with wearable cameras that have sensing capabilities for accessibility (\eg~\cite{goodman2019social, findlater2019fairness, ahmed2018up}). Specifically, we focus on pedestrian detection with wearable cameras for blind users. While this technology can provide equitable access to approaching pedestrians with implications for ensuring one's safety~\cite{branham2017is} or managing social interactions~\cite{ kianpisheh2019face}, its always-on nature raises privacy concerns for parties that may or may not realize that they are being recorded and/or visually interpreted.

The main contribution of this work is that we explore this tension from both perspectives, those of sighted passersby and blind users, taking into account camera visibility and remote versus in-person experience with a working prototype; dimensions lend an opportunity for further insights into prior findings from Profita \etal~\cite{profita2016effect} and Ahmed \etal~\cite{ahmed2018up}. 
More specifically, our first study is a balanced online 2x2 factorial-design experiment. Participants ($N$=206), recruited through Amazon Mechanical Turk, responded to technology attitude and other questions before and after being exposed to videos portraying passerby and user perspectives. Videos varied in terms of the wearable camera (visible head-mounted GoPro versus 'invisible' camera in smart glasses) and in terms of the blind actor (male adult with a white cane versus female older adult with a guide dog), as shown in Figure~\ref{fig:video_desc}.
Our second study is an in-person experience study, where blind participants ($N$=10) wear a working prototype for pedestrian detection and pass by sighted participants ($N$=40). Since sighted participants experience a scenario similar to the passerby perspective video, they are only shown the user perspective video.
Duerden and Witt~\cite{duerden2010impact} indicates that direct experience catalyzes people's perception of technology more than remote experience. We explore how this plays when trying to capture people's perceptions toward assistive wearable cameras for pedestrian detection designed for blind people.

We find that our blind and sighted participants held differing opinions on privacy concerns brought forth by assistive wearable cameras. They displayed a conflict of interest regarding which visual features should be detected by an assistive wearable camera; \eg, blind participants favored estimations of \textit{head pose} and \textit{facial expression} of pedestrians, but sighted passerby participants did not want such an assistive system to detect these features.
We observed that sighted participants changed their opinions by being more reserved on assistive tech with wearable cameras after learning what information it can glean from pedestrians. We noticed that the way bystanders experienced this technology (personally or remotely) seems to relate with their opinions; with in-person participants being more negative though this could be explained by other factors and requires a follow up study to confirm directionality.
Our findings suggest that it is important to understand the conflict of interests between users and bystanders regarding such technology while also confirming that people are amicable toward assistive uses of wearable camera technology.
Our results suggest that addressing the conflict in design choices between potential users and pedestrians would help increase social acceptance of it.
We recommend that other factors, such as the way of exposure to technology, and the awareness of the data being captured and processed by the technology, need to be considered to obtain a clearer understanding of people's perceptions and attitudes.

\section{Related Work}
Our research is built upon prior work concerning the social acceptance of wearable technology as well as face detection and recognition technology for people with visual impairments. We discuss prior work to understand people's perceptions of wearable technology and learn about assistive camera systems designed for people with visual impairments.

\subsection{Social Acceptance of Wearable Devices}
Wearable technology that expects user input often uses gestural interfaces.  As gestural interactions between users and wearable devices can cause social tensions in certain situations, researchers have explored the social acceptance of such technology~\cite{rico2009gestures, montero2010would, rico2010usable, profita2013don, ahlstrom2014you, you2019understanding}.
This social acceptance evaluation focuses primarily on the perspective of users (their perceptions of the technology and their comfort with using such gestures)~\cite{rico2009gestures, montero2010would, rico2010usable}, since users are the only source of input for such wearable devices.
On the other hand, there is prior work evaluating social acceptance of gesture-based wearable technology from the bystander's perspective~\cite{profita2013don, you2019understanding}.  Videos presenting the uses of wearable devices are often provided to participants so that participants can indirectly experience the technology and then complete a questionnaire for social acceptance.
Our work also uses videos that help participants learn about assistive wearable camera technology without being exposed to the technology in person.

Understanding the social implications of wearable technology from a bystander's perspective becomes more popular when concerning wearable camera devices~\cite{mcatamney2006examination, wolf2014lifelogging, denning2014situ, xu2015exploring, profita2016effect, rauschnabel2016augmented, schwind2018need}, as bystanders are often involved with the technology more explicitly than before.
Although wearable camera technology, such as Google Glass, has caused epidemic controversial privacy concerns, prior work shows that people are more positive toward wearable camera technology when aware that such systems are designed for assistive purposes~\cite{profita2016effect}.
Participants in the prior study remotely experienced the technology by viewing videos recorded from the bystander's viewpoint, yet the authors also acknowledge that in-person experiences with the technology (\eg, seeing or interacting with users wearing such cameras) may engender different social implications.
In this paper, we further investigate these unanswered questions regarding the impact of increased awareness of the data used by wearable cameras and in-person versus remote experiences of the cameras on social acceptance.

\subsection{Assistive Wearable Cameras} 
Since wearable computing devices became capable of obtaining visual data via built-in cameras, not only researchers~\cite{fiannaca2014headlock, mcnaney2014exploring, xu2015littlehelper, malu2015personalized, jin2015smart, al2016ebsar, profita2016effect, rauschnabel2016augmented, zolyomi2017technology, sarfraz2017multimodal, stearns2018design} but also companies~\cite{Aira2019, OrCam2019, eSight2019} have discovered the potential of wearable camera technology in augmenting the capabilities of people, especially those with disabilities.
The findings can be categorized into two groups: one that is focused on users of the technology and another that is interested in the opinions of bystanders observing the technology.
Prior work has investigated uses of wearable camera devices in indoor navigation~\cite{fiannaca2014headlock, al2016ebsar}, vision aid~\cite{zolyomi2017technology, stearns2018design}, and face recognition~\cite{jin2015smart, sarfraz2017multimodal, stearns2018automated}, especially for people with visual impairments. Factors for adoption of such technology are evaluated by potential users~\cite{rauschnabel2016augmented, zolyomi2017technology}.
While this work focuses mostly on the user's perspective, there is prior work that uses a bystander's viewpoint to evaluate wearable camera technology~\cite{profita2016effect, ahmed2018up}.
However, little work has been done to combine both of the user's and bystander's perspectives in terms of social acceptance evaluation. We try to note social tensions between users and bystanders by looking at both groups' opinions on wearable cameras for assistive purposes.

\subsection{Face Detection and Recognition for Blind People}
Blind people generally have very limited sources to access information about other people in their daily lives.  This hurdle consequently leads them to difficulty in initiating social interactions where visual cues, such as gaze, are essential for social initiation. Many assistive systems have been developed to help blind people obtain visual cues from other people or recognize their family members and friends~\cite{jayant2011supporting, jin2015smart, kumar2017intelligent, sarfraz2017multimodal, zhao2018face, stearns2018automated, SeeingAI2019, OrCam2019}.
Early work focused more on helping blind and low-vision people take photos of their target~\cite{jayant2011supporting}. From a photo, the proposed system detects human faces and informs of the faces' locations via nonvisual feedback, but this system is not incorporated with face recognition. As computer vision algorithms evolve, face recognition has been employed in assistive systems by using either WiFi-enabled glasses with a camera~\cite{jin2015smart} or a smartphone~\cite{kumar2017intelligent}. The studies claim that their systems help blind people recognize people without evaluation with potential users, raising questions about feasibility and usability of such systems. To address these questions regarding assistive wearable camera systems designed for blind people, we built a working prototype system and evaluated it with blind people.

There is prior work that has explored real use cases of such assistive systems. 
Zhao \etal evaluated an accessibility bot of Facebook messenger on a smartphone with six participants with visual impairments~\cite{zhao2018face}. The accessibility bot not only recognizes names but also provides other information, including facial expressions and attributes, of a person if the person is a friend of the user.  However, the results of participants' actual use cases of this system were very diverse (even among the six participants), making it difficult to extract common patterns in their interactions with the system.  In our paper, we focus on a single use case, \textit{pedestrian detection}, of an assistive wearable camera to limit variations and to understand people's interactions with the wearable technology in depth.
Moreover, another work using multiple cameras (two stationary cameras and two wearable cameras) explored various characteristics --- such as camera placement, field-of-view, and image distortion --- in capturing and detecting people~\cite{stearns2018automated}. The authors highlight that using a wearable camera would be necessary, as it provides egocentric spatial reference to users.
Bystanders (non-users), however, might have different perceptions of such wearable cameras depending on the form or visibility of the cameras.  To further investigate this aspect, we use two wearable cameras (a GoPro camera and a pair of smart glasses) that vary in camera visibility as shown in Figure~\ref{fig:video1}.

\section{Online Survey}
We first explore social acceptance of wearable camera technology for blind people with online sighted participants. Amazon Mechanical Turk was used as it provides easy access to a large pool of people. More than 200 participants in the U.S. answered our questions before and after watching the passerby perspective video and the user perspective video (Figure~\ref{fig:video_desc})\footnote{The questionnaire and videos are available on \url{https://iamlabumd.github.io/chi2020_lee/}}.  We created our online survey through SurveyGizmo and linked it to Amazon Mechanical Turk.

\subsection{Participants}
A total of 206 individuals (74 female, 132 male) from the U.S. were recruited\footnote{\label{note:irb}IRB Protocol Number is \textit{STUDY2019\_00000294}.} via Amazon Mechanical Turk. As in Profita \etal~~\cite{profita2016effect}, we specified the USA-only constraint to minimize potential data variations due to cultural and regulatory differences.  Participants ranged from 19 to 70 in age ($\mu$=35.78, $\sigma$=9.9); only adults, age 18 or older, were allowed to participate in our online survey.
Our online participants tended to be slightly positive toward broader technology ($\mu=1.63$, $\sigma=0.66$)\footnote{\label{note:atdscore1}Responses to attitudes questions are converted into -3 to 3 with -3 indicating the \textit{most negative}, 0 \textit{neutral}, and 3 the \textit{most positive}.} and wearable technology ($\mu=1.16$, $\sigma=1.24$)\footnoteref{note:atdscore1}.

\subsection{Video Stimuli}

\begin{table}[t]
    \centering
    \caption{Video characteristics and reference codes.}
    \begin{tabular}{c c c c c}
        \textit{Camera} & \textit{Blind Actor} & \textit{Assistance} & \textit{Code} \\
        \hline
        GoPro & male & white cane & \textbf{GoPro+cane}\\
        Smart glasses & male & white cane & \textbf{Glasses+cane}\\
        GoPro & female & guide dog & \textbf{GoPro+dog}\\
        Smart glasses & female & guide dog & \textbf{Glasses+dog}\\
    \end{tabular}
    \label{tab:video}
\end{table}

Participants watched two videos that presented assistive wearable camera technology in two different perspectives, respectively: the passerby perspective and the user perspective.
We created the videos using two different wearable devices --- either smart glasses or a GoPro --- and two different sets of actors --- (a female pedestrian and a male blind user with his white cane) or (a male pedestrian and a female blind user with her guide dog). This totaled to four different versions of each video.  Figure~\ref{fig:video_desc} describes the videos, and Table~\ref{tab:video} shows the videos' characteristics and reference codes.

The passerby video presents a blind user wearing a camera device and walking through a corridor. This video shows how blind users wearing such a camera would be observed by pedestrians.
On the other hand, the user video introduces four scenarios using the camera for assistive purposes, focusing more on the user's point of view.
It presents how pedestrians would be captured by the camera as follows:
\begin{description}
    \setlength\itemsep{0em}
    \item [Scenario 1] The wearable camera detects the pedestrian's presence and estimates the distance between the user and pedestrian as well as the head pose of the pedestrian.
    \item [Scenario 2] The wearable camera detects simple visual features of the pedestrian, such as age and gender, as well as the pedestrian's distance and head pose.
    \item [Scenario 3] The wearable camera not only estimates the distance and head pose but also detects more visual features, such as age, gender, ethnicity, and hair color.
    \item [Scenario 4] The wearable camera recognizes the pedestrian and estimates the distance and head pose.
\end{description}

\subsection{Procedures}
On the instructions page, we first mentioned that our online survey contained secret questions for checking participation and determining compensation to encourage participants to read and answer all the questions thoroughly; it was adapted from the performance-based payment approach~\cite{ho2015incentivizing}.
After that, the online survey opened with demographic questions and several questions about attitudes toward technologies. Once participants completed this questionnaire, the survey platform randomly assigned a pair of videos showing a different set of actors and either of the wearable cameras.
On the survey, we described that the devices assisted blind people.

There were two secret questions (one after the passerby video and another after the user video) that we used to check whether online participants watched the videos; we marked a response as invalid and did not use it for our analysis if a participant had a wrong answer to either of these questions.
The secret questions asked about what the blind user was holding in the passerby video and what features the prototype system detected from the pedestrian in the user video.

After watching the passerby video, participants were asked to answer questions about social acceptance of assistive wearable camera technology. Then, they watched the user video and were asked if they would be okay with each of the scenarios. Once they responded to the scenario questions, they were asked to answer some of the social acceptance questions, which were the same as the ones asked after the passerby video.
These questions were intentionally asked again to study changes in the participants' perceptions of the technology after the user video that informed them of what sort of information the system collected from pedestrians and used for assistive purposes.
The back button on the online survey was disabled to prevent participants from altering their answers to previous questions. 
Participants were compensated \$2 for their participation as long as they completed the survey.

\subsubsection{Questionnaire}
A questionnaire about social acceptance toward assistive wearable camera technology consists of questions that are grouped into eight categories:
\begin{description}
    \setlength\itemsep{0em}
    \item [OkayAssistUse] Okay to see blind people using the technology in public.
    \item [OkayAnyUse] Okay to see any people using the technology in public.
    \item [PrivacyConcerns] Privacy issues with blind people using the technology in public.
    \item [UncomfortableAssistUse] Uncomfortable to see blind people using the technology in public.
    \item [OkayRecording] Okay with the camera recording photos/videos for improvement.
    \item [OkaySingleUse] Okay with the camera using photos/videos for one-time detection.
    \item [UncomfortableRecording] Uncomfortable with the camera recording photos/videos.
    \item [UncomfortableSingleUse] Uncomfortable with the camera using photos/videos for one-time detection.
\end{description}

Questions under four categories (\textit{OkayAssistUse, OkayAnyUse, PrivacyConcerns, UncomfortableAssistUse}) were asked after the passerby video and the user video, respectively.
We created questions based on prior questions asking about social acceptance of wearable technology and attitudes toward technology~\cite{kelly2016wear, rosen2013media}.
For some aspects of social acceptance, we asked participants about their \textit{acceptance} (\ie, Okay-) and \textit{discomfort} (\ie, Uncomfortable-), as people might accept the technology but feel uncomfortable about it, or vice versa.

In addition, after the user video, participants answered questions about the four different uses of such assistive technology as well as one additional use (\textit{Recognition of any people}).  Following are the descriptions of the five questions:
\begin{description}
    \setlength\itemsep{0em}
    \item [Scenario 1 (presence, distance)] Okay with the camera detecting my presence and distance between the user and me.
    \item [Scenario 2 (basic visual features)] Okay with the camera detecting my basic visual attributes, such as age and gender.
    \item [Scenario 3 (more visual features)] Okay with the camera detecting more of my visual attributes, such as age, gender, hair color, facial expression, ethnicity, and actions.
    \item [Scenario 4 (recognition)] Okay with the camera recognizing me (i.e., my name) when I know the blind user.
    \item [Recognition of any people] Okay with the camera recognizing me (i.e., my name) when I don't know the blind user.
\end{description}

Participants were asked to respond all the questions above with a 7-point Likert scale: \textit{Strongly disagree, Moderately disagree, Slightly disagree, Neutral, Slightly agree, Moderately agree, Strongly agree}.


\subsection{Results}
Among the total 206 responses, 22 incorrectly answered to our secret question(s) and were thus excluded from our analysis. As a result, we obtained 45, 47, 45, and 47 valid responses from participants who watched GoPro+cane, Glasses+cane, GoPro+dog, and Glasses+dog video pairs, respectively.  Only the first 45 responses for each video pair were used for our analysis; hence, our analysis used 90 responses from each camera form (Glasses/GoPro).
For analysis, we converted their responses from the 7-point Likert scale into numeric values ($-3$ to $3$): $-3$ for \textit{Strongly disagree}, $0$ for \textit{Neutral}, and $3$ for \textit{Strongly agree}.

\begin{figure}[t]
    \centering
    \includegraphics[width=0.47\textwidth]{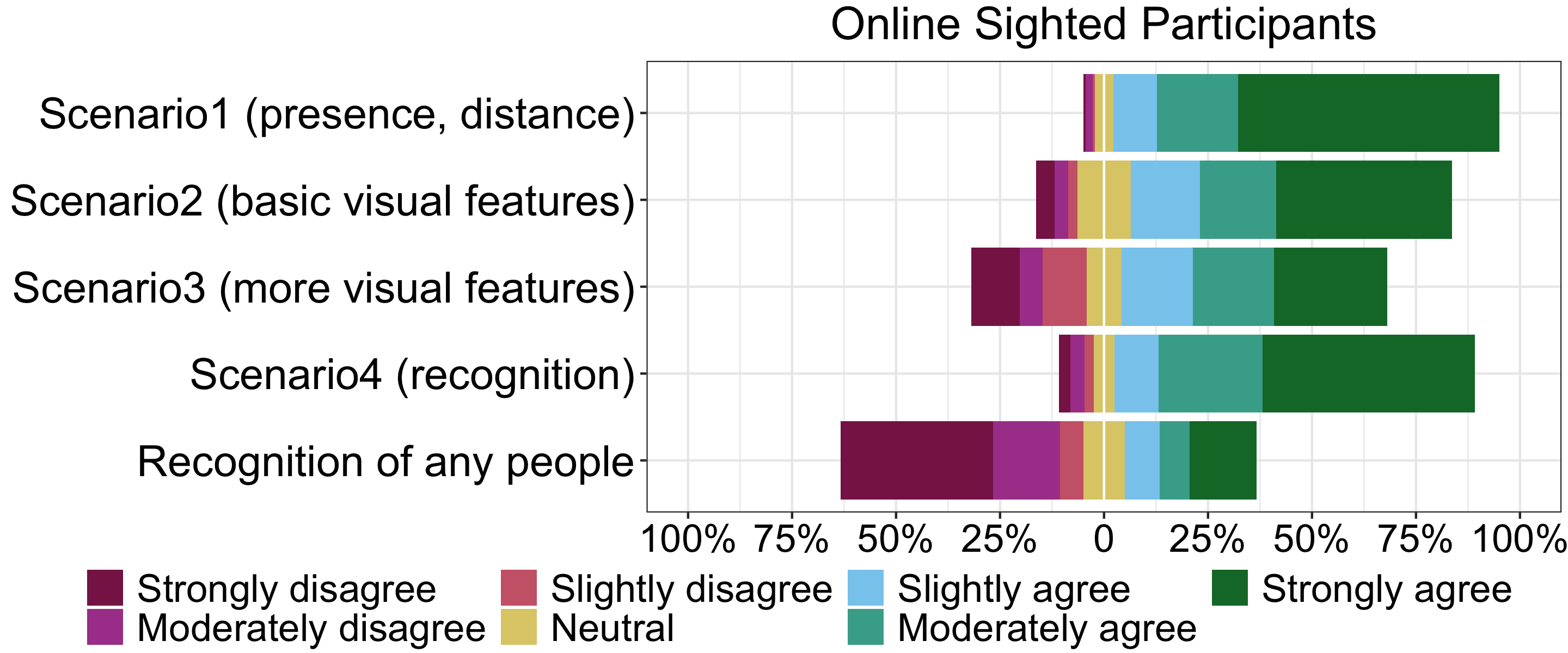}
    \caption{Online sighted participants' opinions on use cases of assistive wearable cameras. Participants were asked whether they would be okay with each of the use cases.}
    \label{fig:remote_scenarios}
\end{figure}

\subsubsection{Perspectives of online sighted participants}
Using all responses from the Glasses and GoPro groups, we first examined participants' perceptions of assistive wearable camera technology based on their responses to the social acceptance questions answered after the passerby video.
In general, our participants tended to be positive toward the assistive wearable technology for all the eight categories.
Online sighted participants tended to agree with the positive statements about the technology for assistive purposes, such as \textit{OkayAssistUse} ($\mu=2.2$, $\sigma=1.1$), \textit{OkayRecording} ($\mu=1.5$, $\sigma=1.6$), and \textit{OkaySingleUse} ($\mu=1.8$, $\sigma=1.5$).  They generally showed agreement even on the statement about any person using the technology: \textit{OkayAnyUse} ($\mu=1.0$, $\sigma=1.6$).
We also noticed that our online participants tended to disagree on the negative statements: \textit{PrivacyConcerns} ($\mu=-0.4$, $\sigma=1.8$), \textit{UncomfortableAssistUse} ($\mu=-1.4$, $\sigma=1.7$), \textit{UncomfortableRecording} ($\mu=-0.1$, $\sigma=2.1$), and \textit{UncomfortableSingleUse} ($\mu=-1.3$, $\sigma=1.9$).

Figure~\ref{fig:remote_scenarios} visualizes their responses to the scenario questions.
We observed that the less features the assistive camera system detected from pedestrians, the more positive participants tended to be about the system. Furthermore, more than 85\% of the participants answered that they would be okay with the system recognizing them if they knew the users. As for the question about \textit{Recognition of any people}, more than 30\% of the participants were amicable to the system recognizing people even if they did not know users, but more than half the participants disagreed to this.


\subsubsection{Camera visibility}

\begin{figure}[!t]
    \centering
    \begin{subfigure}[t]{0.15\textwidth}
        \centering
        \includegraphics[width=\textwidth]{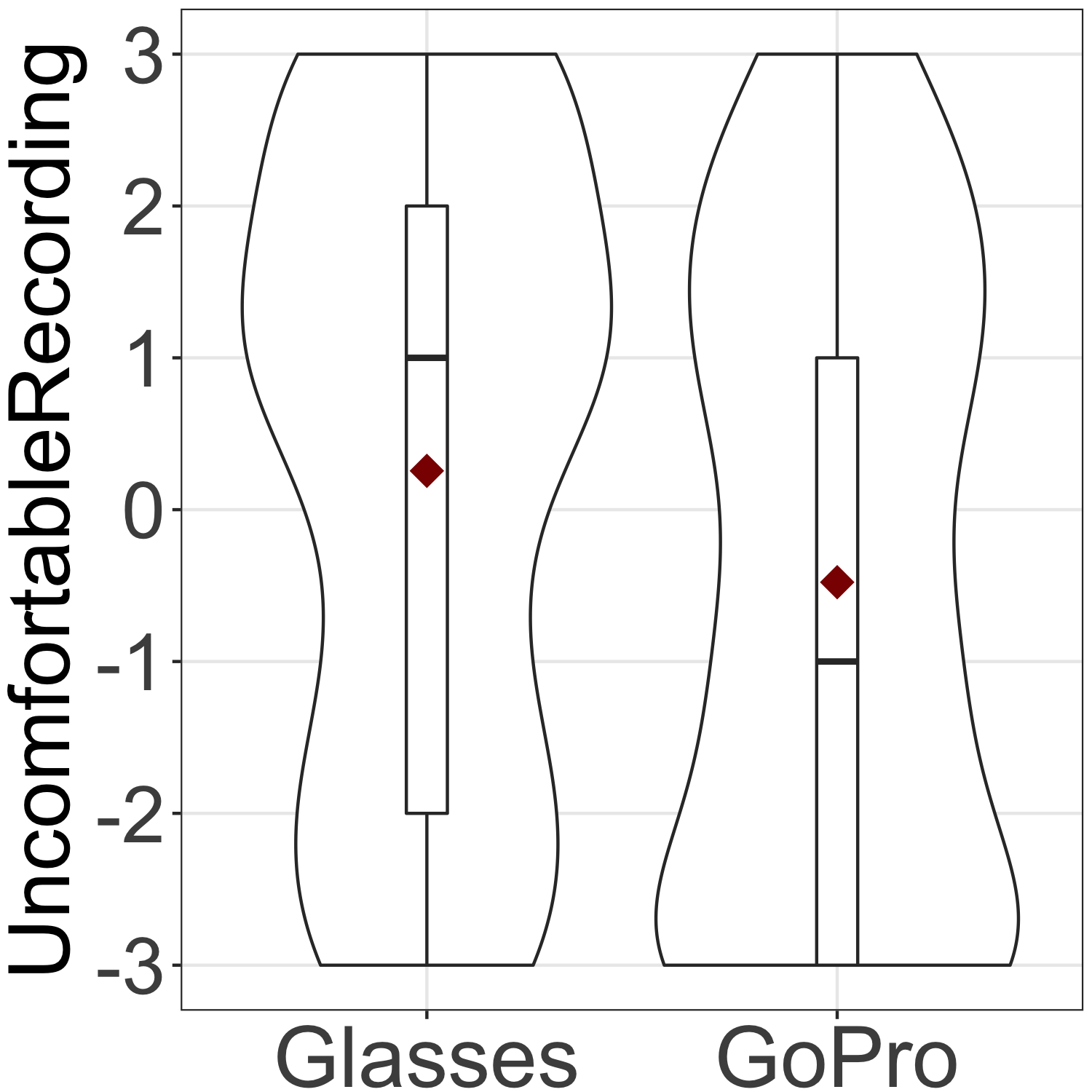}
        \caption{Participants in Glasses felt more uncomfortable about camera recording than participants in GoPro.}
        \label{fig:online_pre_v1q6}
    \end{subfigure}%
    \hspace{0.44em}
    \begin{subfigure}[t]{0.15\textwidth}
        \centering
        \includegraphics[width=\textwidth]{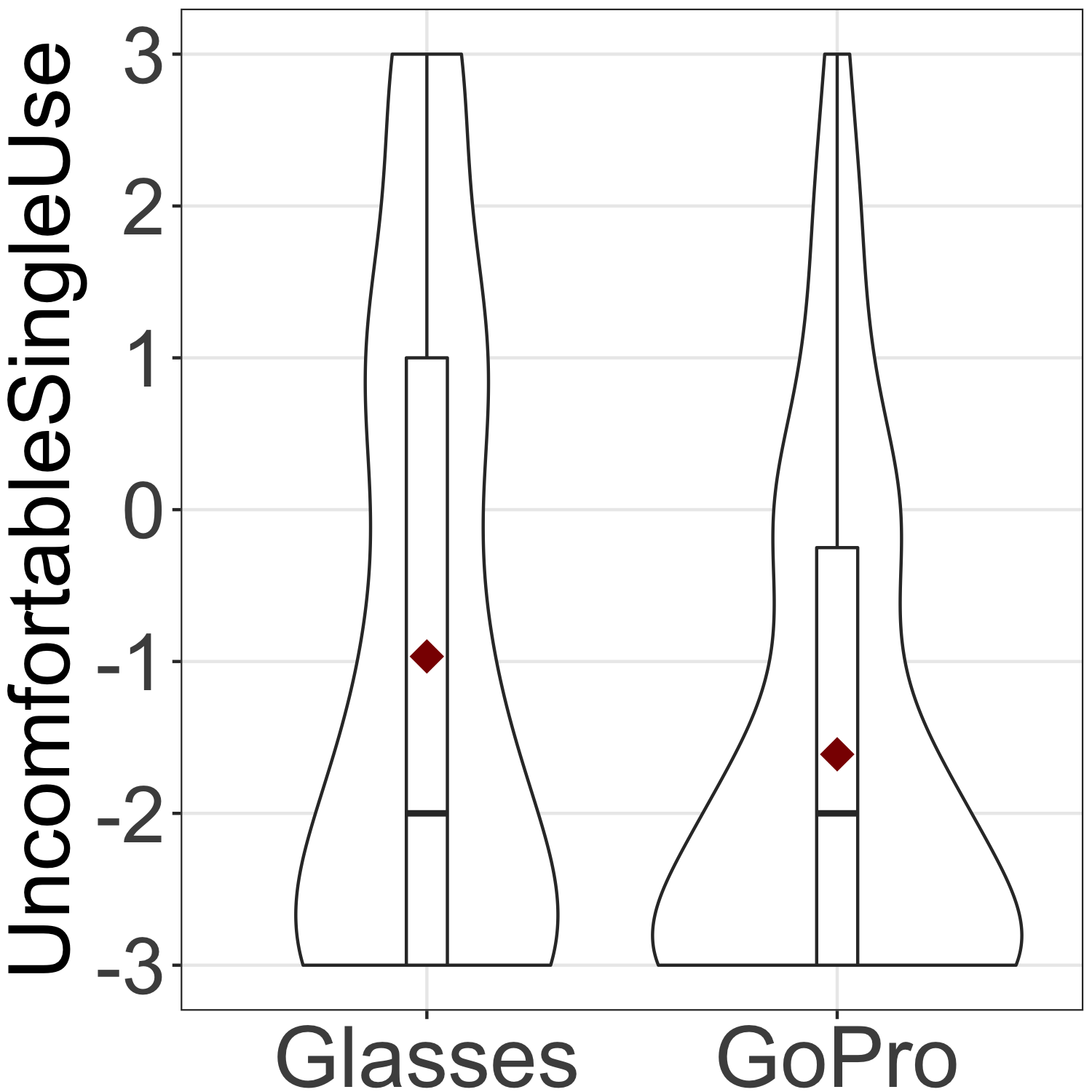}
        \caption{Participants in Glasses felt more uncomfortable about single data use than participants in GoPro.}
        \label{fig:online_pre_v1q8}
    \end{subfigure}%
    \hspace{0.44em}
    \begin{subfigure}[t]{0.15\textwidth}
        \centering
        \includegraphics[width=\textwidth]{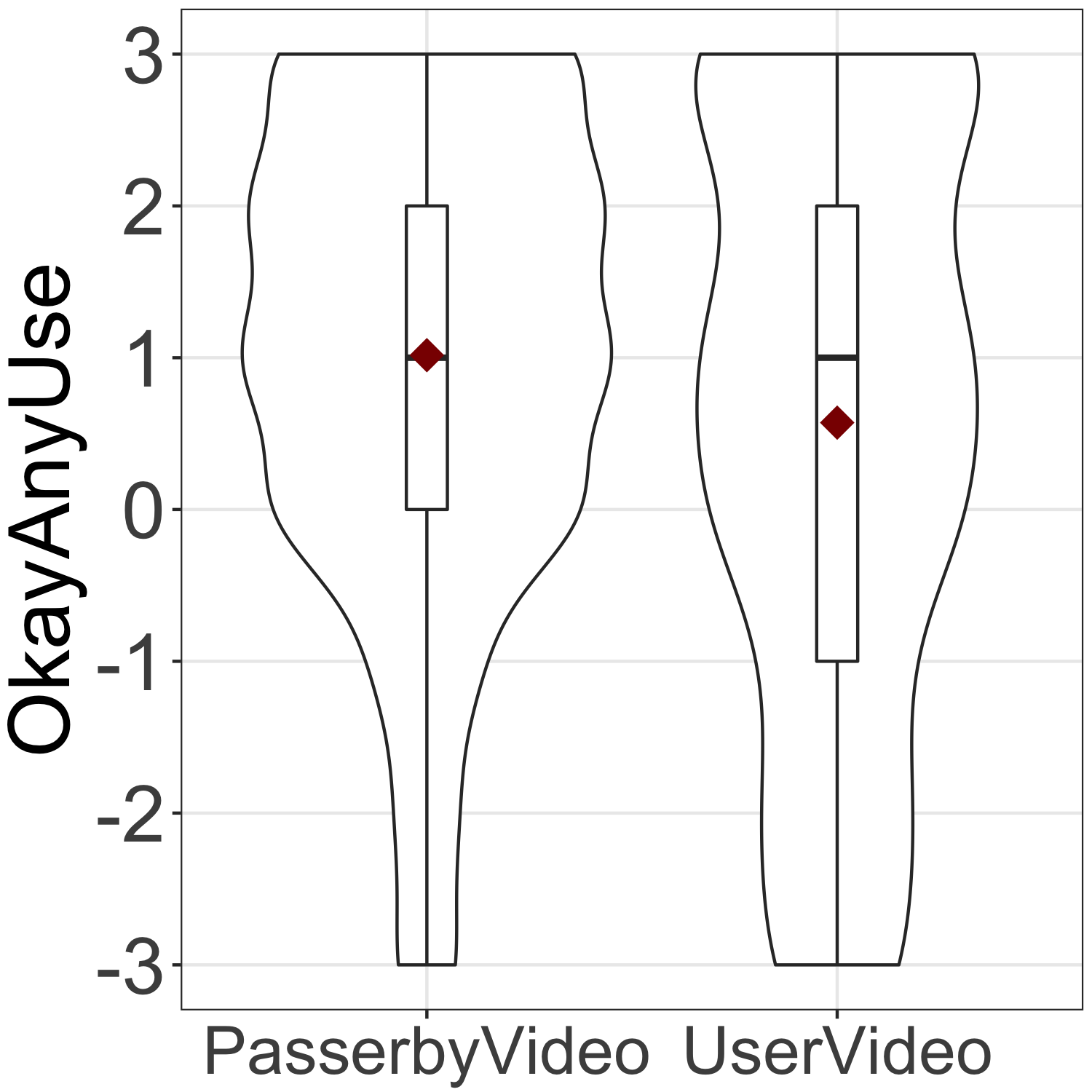}
        \caption{Participants became more negative about any people using the technology after watching the user perspective video.}
        \label{fig:online_any_before_after}
    \end{subfigure}%
    \caption{Comparisons of online participants' responses.}
    \label{fig:online_boxes}
\end{figure}

Comparing responses from the Glasses and GoPro, we then sought to see whether camera visibility affects people's perceptions of the technology.
We first counted the number of participants who agreed that it was easy to notice a camera in the passerby video; this question was asked right after they watched the passerby perspective video.
Of the 90 participants in GoPro, 78 (87\%) agreed that it was easy to notice the camera while only 14 (16\%) of the 90 participants in Glasses agreed on it.  Using the Mann-Whitney-Wilcoxon test for significance testing and the Rosenthal's formula for effect size calculation, we found that the difference in camera visibility between the cameras was high and significant ($p<0.01$, $r=0.715$).

We further analyzed their responses to the social acceptance questions asked after the passerby video to see effects of camera visibility on their perceptions of the technology.
Figure~\ref{fig:online_pre_v1q6} and \ref{fig:online_pre_v1q8} demonstrate that participants in Glasses agreed to feeling more uncomfortable about a wearable camera that either records them ($p<0.05$, $r=0.176$) or uses images/videos for detection without saving ($p<0.05$, $r=0.152$) than participants in GoPro were.


\subsubsection{Awareness of attributes detected by technology}
Social acceptance of wearable camera technology has been primarily evaluated from a bystander's view point, meaning that potential bystanders may not know about what data the technology will take to accomplish its objectives.  To evaluate effects of awareness of the data processed by the wearable technology on social acceptance, we examined participants' responses to the social acceptance questions repetitively asked before and after the user perspective video.  Note that the user video informed of the wearable camera's function of capturing visual attributes of pedestrians.  For this analysis, all the responses from the Glasses and GoPro groups were merged and used, and the Wilcoxon signed-rank test was used to compare their responses before and after the user video.


\begin{figure}[t]
    \centering
    \includegraphics[width=0.45\textwidth]{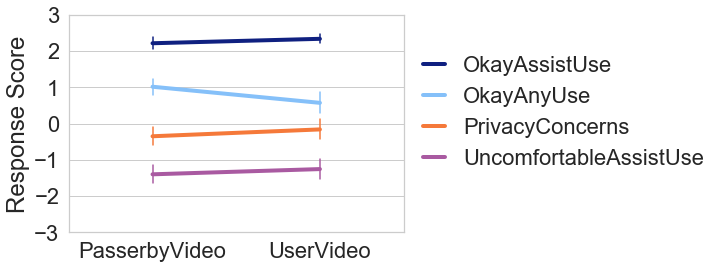}
    \caption{Online sighted participants' responses to the social acceptance questions before and after the user perspective video.  Responses are converted to numeric values ($-3$ to $3$): $-3$=\textit{Strongly disagree}, $0$=\textit{Neutral}, $3$=\textit{Strongly agree}.}
    \label{fig:online_social_acceptance}
\end{figure}

As illustrated in Figure~\ref{fig:online_social_acceptance}, participants generally tended to be positive toward wearable camera technology if it was designed for assistive purpose after watching the passerby video and the user video.
However, we observed that more participants tended to become uncomfortable seeing people using such a wearable camera device in public spaces, even if it was for assistive purpose, after the user video  (\textit{UncomfortableAssistUse}).  The number of participants who agreed to feeling uncomfortable increased after the user video (from 38 to 50 out of the 180 participants), but the difference was not statistically significant.
For one category, \textit{OkayAnyUse}, participants became more negative toward it after they watched the user video ($p<0.01$, $r=0.265$); Figure~\ref{fig:online_any_before_after} shows their responses before and after the user video.
In particular, one online participant strongly agreed that any people can use such technology by saying ``people can wear whatever they want'' before the user video. However, after learning about visual features estimated by the technology, the participant switched to strongly disagreeing with this category and commented ``I would feel like they're invading my privacy.''

Lastly, we questioned whether camera visibility affects the participants' perceptions of the technology before and after they are aware of data processed by the technology.
Participants in Glasses and GoPro agreed to change their perceptions of the technology after watching the user perspective video, but participants in Glasses ($\mu=1.41$, $\sigma=1.54$) agreed more to change their perceptions than did participants in GoPro ($\mu=0.73$, $\sigma=1.87$): $p<0.01$, $r=0.182$.
Using ANCOVA, we checked if there was a significant difference in perception changes due to camera visibility. The result indicates that there is a strong interaction between the camera visibility and the perception changes; \ie, the camera visibility affected the changes in their opinions on social acceptance of wearable technology before and after the user video ($p<0.05$).


\section{GlAccess: A Pedestrian Detection Prototype}
\begin{figure}[t]
    \centering
    \includegraphics[width=0.48\textwidth]{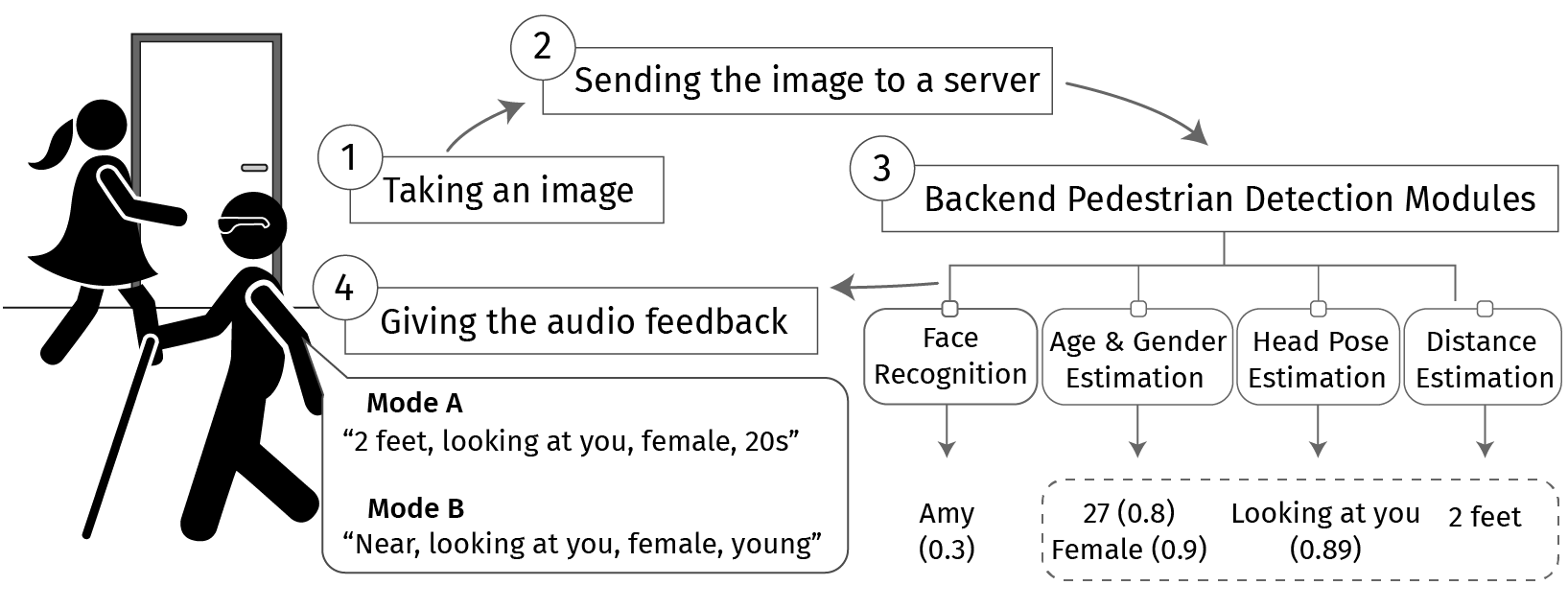}
    \caption{Diagram of data processing of \textit{GlAccess}.}
    \label{fig:system_diagram}
\end{figure}

We further wanted to know how people would perceive assistive wearable cameras if they were exposed to such technology in person. For this evaluation, we built a working prototype system using a wearable camera device to conduct an in-situ user study.
The prototype system is designed to help blind people access visual information about pedestrians around them. We call our prototype system \textit{GlAccess} (Glasses + Access), which estimates visual attributes of people who pass by our system users.
\textit{GlAccess} consists of a server and a client. The client is an app running on a pair of smart glasses that communicates with the server to obtain visual features of a pedestrian as described in Figure~\ref{fig:system_diagram}.

\subsection{Server}
The server is responsible for detecting visual features of and recognizing a pedestrian captured in photos sent by the client.  Once the server receives a photo from the client, it tries to estimate visual features --- such as name, age, gender, and head pose --- of a person in the photo.  Following is a list of capabilities of the server: face detection and recognition, and age, gender, head pose, and distance estimations.

\subsection{Client}
The client is responsible for delivering feedback to users based on the image analysis from the server.  It continuously requests the server to analyze one image at a time.  Once the system detects a pedestrian, it generates a text message and speaks it by using text-to-speech.  The client remains silent if no one is detected.
To prevent message delivery from being interrupted by new responses, the client will ignore the new message if it is in the middle of message delivery of a previous message.


\subsubsection{Feedback Message}
The client has two different modes, \textit{Mode A} and \textit{Mode B}.
The system composes the messages in order of distance, head pose, (name or gender \& age), and position; \eg, ``15 feet, looking at you, male, 30s, on the left.''  The gender and age estimations are provided to users if the system is not confident of the name estimation (\ie, face recognition).
Each mode has a different way of creating a feedback message for distance and age; but, the other features (head pose, name, gender, position) are delivered same way in both Mode A and B.
If the confidence score\footnote{0 $\le$ confidence score $\le$ 1; the higher, the more confident.} of each estimation is below a certain threshold, the system does not deliver that estimation.
Based on our pilot testing, we empirically set two different thresholds: $0.5$ for gender, age, and name estimations, and $0.7$ for head pose estimation.
Note that the age and head pose estimations are binary classifications, and the distance and position estimations do not have confidence scores as these estimations are proportionally computed with the size of a detected face.

\begin{description}
    \setlength\itemsep{0em}
    \item [Mode A:] Quantitative feedback mode.  It uses the foot metric for distance estimation and approximation for age estimation ---  \eg, ``2 feet, looking at you, female, 20s.''
    \item [Mode B:] Concise feedback mode.  It simply says whether a pedestrian is detected to be \textit{near} (< 15 feet) or \textit{far}.  Age is estimated as \textit{young} (age < 35), \textit{middle-age} (35 $\le$ age < 55), or \textit{older} --- \eg, ``Near, looking at you, female, young.''
\end{description}

\subsection{Implementation}
The server application was written in Python with the Flask framework and run on a GPU-powered Ubuntu 18.04 Linux system.
For face detection and alignment, we employed Insightface~\cite{deng2018arcface} built upon the Multi-task Cascaded Convolutional Networks (MTCNNs)~\cite{zhang2016joint} that detects faces.  This model was composed of several models that not only detect face embedding (512-dimension data) but also estimate age and gender of a detected face.  Using the face embedding data, we trained a SVM classifier to recognize five of our lab members; blind participants interacted with them during user study (see next section).
Moreover, we fine-tuned\footnote{The SGD optimizer was used with the following hyperparameters: learning rate=0.005, batch size=8, epochs=100.} the face embedding model on a dataset for head pose estimation; the dataset included the head pose database~\cite{gourier2004estimating} and our head pose images collected using the smart glasses.  The head pose estimation model simply checked whether or not a pedestrian was looking at a user.
The distance between a pedestrian and a user was estimated based on the height of a detected face. The distance estimation function was defined with pairs of the actual distance and the face height from our dataset.
Since our user study was conducted in the U.S., we used \textit{foot} for distance.

The client system was implemented as an Android app for smart glasses, Vuzix Blade~\cite{VuzixBlade}. The smart glasses ran on Android OS (5.1) and were connected to the server through Wi-Fi. Since the glasses did not have a speaker, a Bluetooth headset with earbuds was used so that users could hear audio messages; using a bone conduction headset was not feasible as its body conflicted with the glasses.
During the development, we observed that it usually took less than one second to send a request to and receive a response from the server.

\section{In-situ Study}

The goal of our in-situ study is to understand conflicts of interest in such assistive technology between blind users and pedestrians and to further evaluate their opinions on the technology by comparing results from our online survey.

\subsection{Participants}
We recruited\footnoteref{note:irb} a total of 50 participants: 10 blind participants and 40 sighted passerby participants.

\begin{table}[t]
 \setlength{\tabcolsep}{3pt}
 \renewcommand{\arraystretch}{1}
    \centering
    \caption{Demographic information about blind participants.}
    \begin{tabular}{ c c c c c c c }
    \textit{PID} & \textit{Gender} & \textit{Age} & \textit{Blindness} & \textit{\makecell{Light\\Persept.}} & \textit{Onset} & \textit{\makecell{Wearable\\Camera}} \\
    \hline
     P1 & F & 71 & Totally & N & 63 & Y \\
     P2 & M & 66 & Totally & N & birth & Y \\
     P3 & M & 59 & Totally & N & birth & N \\
     P4 & F & 66 & Totally & Y & 17 & N \\
     P5 & F & 65 & Totally & N & birth & N \\
     P6 & F & 73 & Totally & N & birth & N \\
     P7 & F & 49 & Totally & N & 35 & N \\
     P8 & M & 60 & Legally & Y & 4 & Y \\
     P9 & M & 56 & Totally & N & 30 & N \\
    P10 & M & 71 & Totally & Y  & 30 & Y\\
    \end{tabular}
    \label{tab:blind_demographic}
\end{table}

\textit{Blind participants}.
Their ages ranged from 49 to 73 years ($\mu$=63.6, $\sigma$=7.6).  As described in Table~\ref{tab:blind_demographic}, four (P1, P2, P8, P10) of them had prior experience with assistive wearable cameras, such as Aira~\cite{Aira2019}, but many of them mentioned that they only used the technology a few times for trial purposes (a few times a month).
Their responses to pre-study questions indicate that our blind participants tended to be positive about general technology ($\mu=2.02$, $\sigma=0.42$)\footnoteref{note:atdscore1} and wearable technology ($\mu=1.89$, $\sigma=0.90$)\footnoteref{note:atdscore1}.

\textit{Sighted participants.}
For each user study, we recruited four sighted passerby participants on our campus; as a result, there were 40 sighted passerby participants in total: 6 faculty/staff, 25 students, and 9 visitors.  Their ages ranged from 19 to 39 years ($\mu$=25.3, $\sigma$=4.8).
Based on their responses to general attitude questions in the post-study sessions, we suspect that they were slightly positive about wearable technology ($\mu=1.08$, $\sigma=0.95$)\footnoteref{note:atdscore1}; there was no pre-study session for the sighted.


\subsection{Study Procedures}

\begin{figure}[!t]
    \centering
    \includegraphics[width=0.45\textwidth]{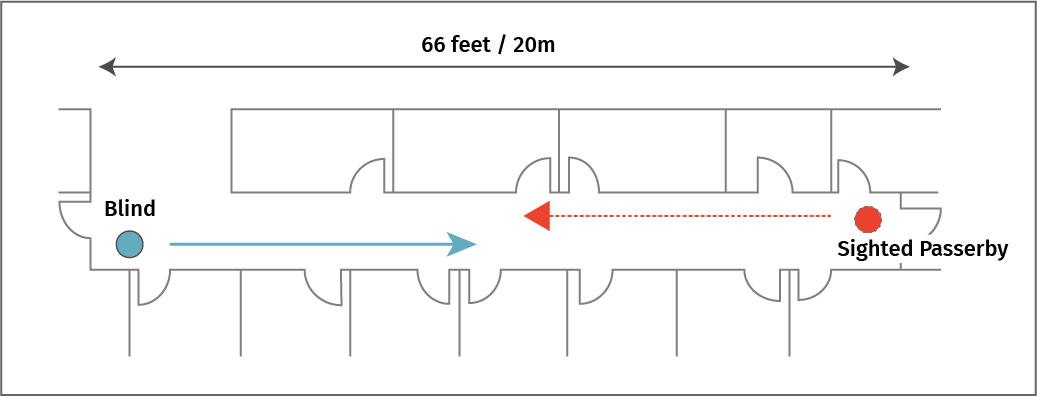}
    \caption{Starting points and walking directions of blind participants and sighted passerby participants in a 66-feet (20-meter) corridor.  A blind participant wearing \textit{GlAccess} and a sighted passerby participant walk toward each other.}
    \label{fig:study_map}
\end{figure}

There were four sessions (\textit{pre-study}, \textit{practice}, \textit{main}, \textit{post-study}) for blind participants and two sessions (\textit{main}, \textit{post-study}) for sighted passerby participants. In the main session, blind users and sighted passerby walked through a corridor and passed by each other as shown in Figure~\ref{fig:study_map}; there were one blind participant and one sighted passerby in the corridor at a time.
Each blind participant passed by eight sighted passerby people: four people (two sighted participants and two of our lab members) for each mode evaluation (Mode A/B).
In each mode evaluation, we also asked first two sighted people to look as they would normally and the other two people to look at their phones while walking through the corridor.

\subsubsection{Blind participants}

\begin{table*}[h]
    \centering
    \caption{Categories and descriptions of social acceptance questions for blind participants, and mean (std.) of scores of their responses to the questions.  Responses are converted to numeric values ($-3$ to $3$): $-3$=\textit{Strongly disagree}, $3$=\textit{Strongly agree}.}
    \begin{tabular}{m{0.23\textwidth} m{0.57\textwidth} m{0.1\textwidth}}
        \textit{Category} & \textit{Description} & \textit{Avg. Score}\\
        \hline
        \textbf{OkayUseInPublic} & Okay to wear the smart glasses in public & 2.6 (0.7) \\
        \textbf{OkayAnyUse} & Okay with any people using the technology in public & 2.2 (1.0) \\
        \textbf{ComfortableUse} & Feel comfortable using the smart glasses & 2.5 (0.5) \\
        \textbf{FeelAccomplished} & Feel more accomplished due to this wearable technology & 1.6 (1.5) \\
        \textbf{HelpInteractWithKnown} & Face recognition helps initiate social interactions with known people & 2.1 (1.0) \\
        \textbf{HelpInteractWithUnknown} & Feature detection helps initiate social interactions with unknown people & 1.4 (1.6) \\
        \textbf{OkayRecording} & Willing to provide many photos/videos of known people if they agree & 2.1 (1.1) \\
        \textbf{PrivacyConcerns} & Privacy issues with blind people using the technology in public & $-1.0$ (2.0) \\
        \textbf{DifficultFaceRecognition} & Face recognition using a wearable camera is difficult & $-2.0$ (1.3) \\
        \textbf{DifficultFeatureDetection} & Feature detection using a wearable camera is difficult & $-2.0$ (0.9) \\
    \end{tabular}
    \label{tab:blind_social_acceptance}
\end{table*}

On average, the user study took two hours. Once blind participants arrived, they first met with our lab members who later participated in their user study as passerby so that the blind participants would recognize the lab members.
Then, we began the pre-study session by obtaining their demographic information, attitudes toward general technology and wearable technology, and prior experience with wearable cameras.  After that, the blind participants wore smart glasses and started using \textit{GlAccess} to detect and recognize a person.  During this practice, blind participants tried Mode A and Mode B to familiarize themselves with each of the modes. Once blind participants became familiar with our system, we started the main study session with sighted participants and our lab members.  
During this session, blind participants used \textit{GlAccess} and were asked to perform the following instruction --- ``When the system detects that a person is nearby, ask the person about the closest office number.''
Using each feedback mode, blind participants walked in the corridor four times to detect or recognize four different people, including two recruited passerby participants who were unknown to our system and two of our lab members who were known to our system, and performed the instruction given to them in the beginning of the session.
After that, they answered questions about their experience and perceptions of the system.  Table~\ref{tab:blind_social_acceptance} describes ten categories of the questions.


\subsubsection{Sighted passerby participants}
Sighted passerby participants were asked to simply walk through the corridor. Before the study, we told them that there would be a camera recording when they walk in the corridor, but did not inform them of a wearable camera device.
Once sighted participants finished walking in the corridor (\ie, the walking activity), they completed our post-study questionnaire, which included the same questions of our online survey; but excluded the passerby video since they were personally engaged in this video's situation during the user study.

\subsection{Results}

\subsubsection{System performance measured with log data}

We first measured the performance of visual feature estimations of our prototype system to check if feedback from the system properly informed blind participants of the pedestrians' attributes. For this measurement, we manually annotated images that our system used to generate and deliver feedback messages about pedestrians to blind participants.
The system achieved, on average, f-score of 0.772, 0.778, 0.775, and 1.0, on estimations of name, gender, head pose, and position, respectively.
The position estimation achieved the maximum performance, meaning that,  whenever the prototype system detected a face, the detected face was a pedestrian's and thus estimated the face position, correctly.

\subsubsection{Perspectives of blind users}
In the post-study session, we observed that some blind participants only thought that the gender estimation was somewhat problematic; although the f-scores of the name and head pose estimations are similar to the gender estimation's.
It seems that this discrepancy may have been due to different levels of attention to visual features, since we learned that some blind participants focused on gender estimation to appropriately address unknown pedestrians using either \textit{sir} or \textit{ma'am} when initiating a conversation; \eg, P6 said, ``(when the system detected a man) I said `excuse me, sir' and it's a woman, ... that's sort of embarrassing.''


Furthermore, we noticed that blind participants tended to be moderately positive about assistive wearable cameras according to their responses to the questions described in Table~\ref{tab:blind_social_acceptance}.
They tended to be positive about using such systems in public (\textit{OkayUseInPublic}) and agreed that it could help them interact with not only known people but also unknown people (\textit{HelpInteractWithKnown}, \textit{HelpInteractWithUnknown}).
Blind participants ($\mu=-1.0$, $\sigma=2.0$) tended to disagree more with there being privacy concerns about the assistive technology than sighted participants (online (Glasses): $\mu=-0.1$, $\sigma=1.8$, in-situ: $\mu=0.2$, $\sigma=1.8$).
Also, they generally agreed to willingly provide more photos/videos of known people to improve recognition system, so long as those known people agree.

\begin{figure*}[t]
    \centering
    \begin{subfigure}[t]{0.155\textwidth}
        \centering
        \includegraphics[width=\textwidth]{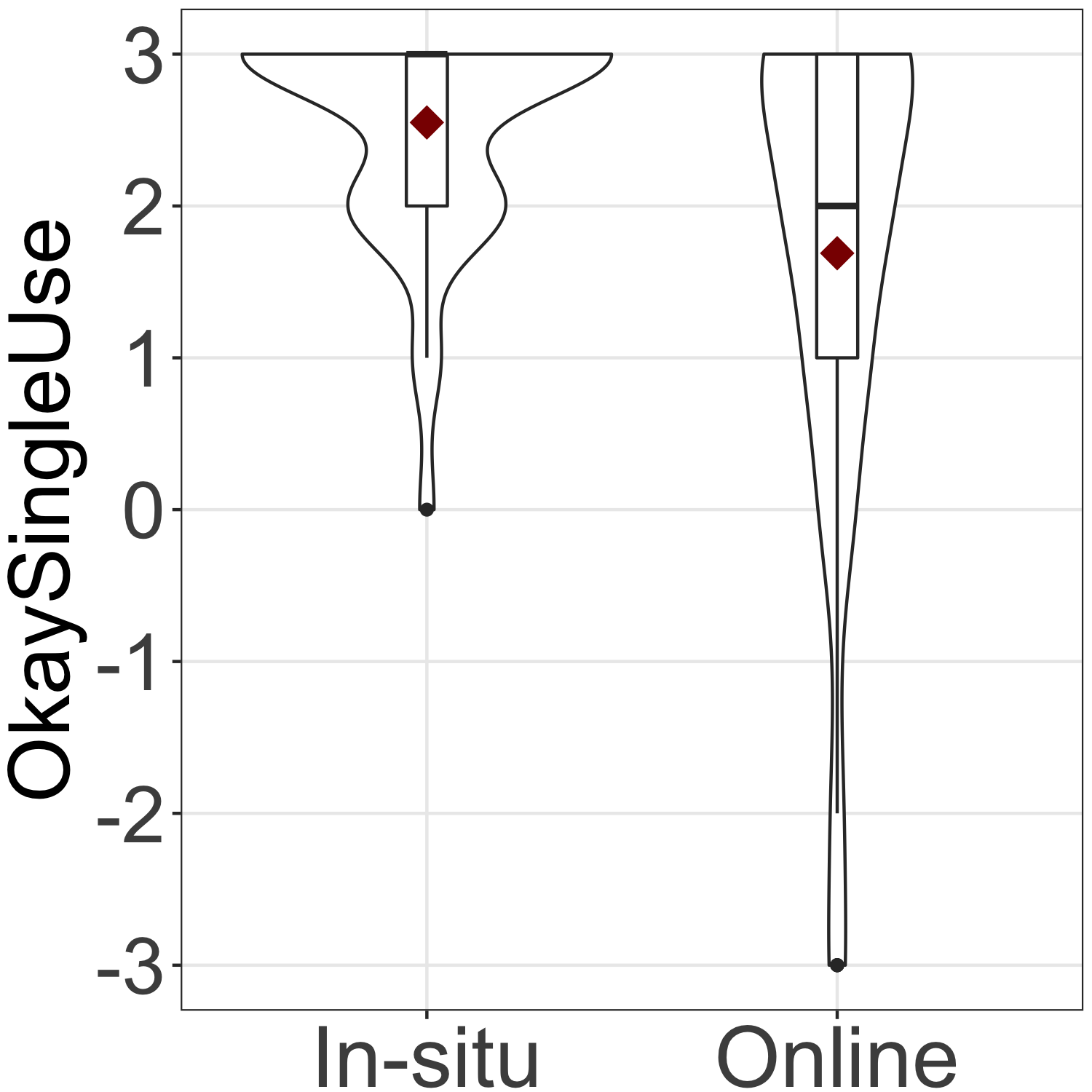}
        \caption{The in-situ sighted were more positive toward single data use than the online sighted.}
        \label{fig:diff_v1q7}
    \end{subfigure}%
    \hspace{0.45em}
    \begin{subfigure}[t]{0.155\textwidth}
        \centering
        \includegraphics[width=\textwidth]{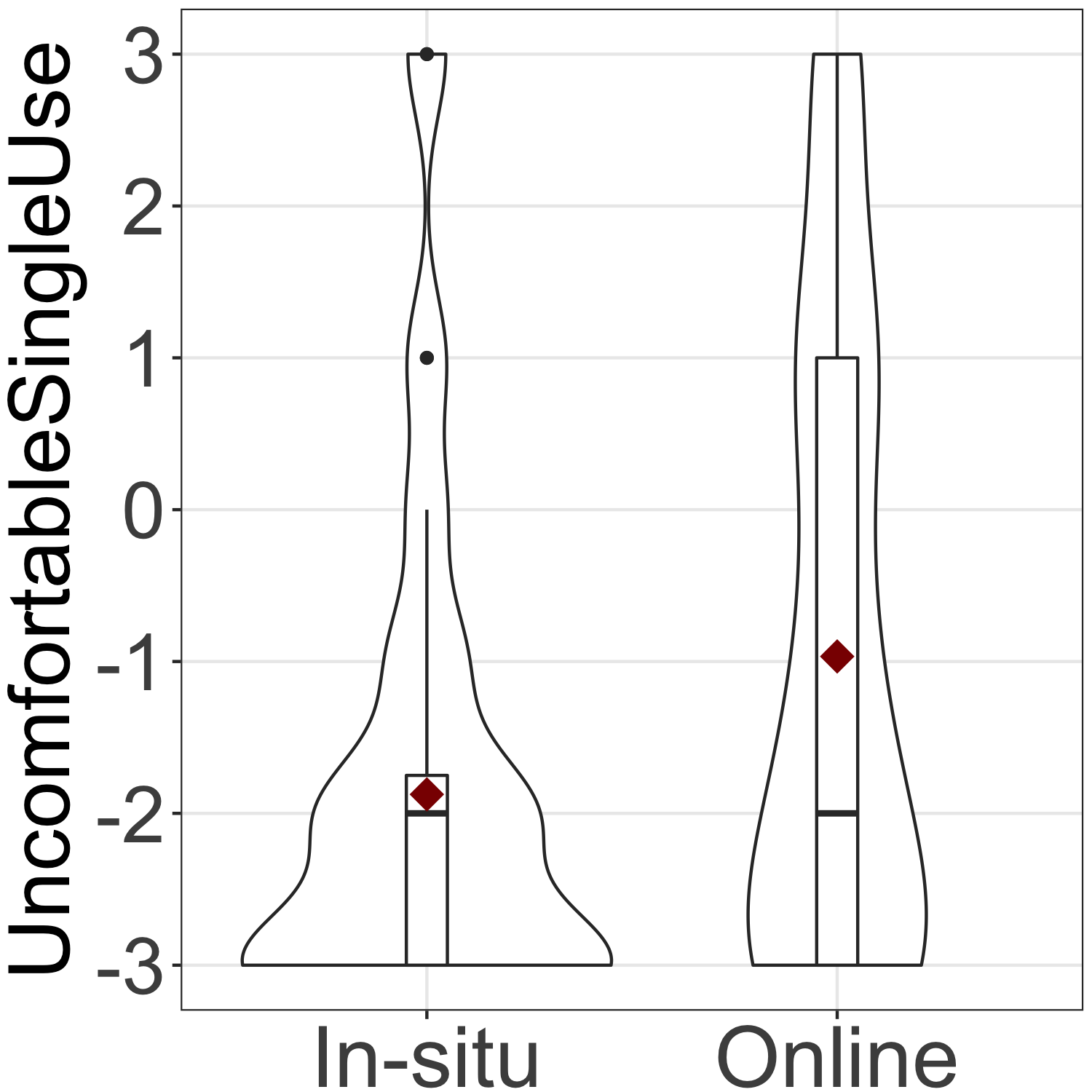}
        \caption{The in-situ sighted felt less uncomfortable toward single data use than the online sighted.}
        \label{fig:diff_v1q8}
    \end{subfigure}%
    \hspace{0.45em}
    \begin{subfigure}[t]{0.155\textwidth}
        \centering
        \includegraphics[width=\textwidth]{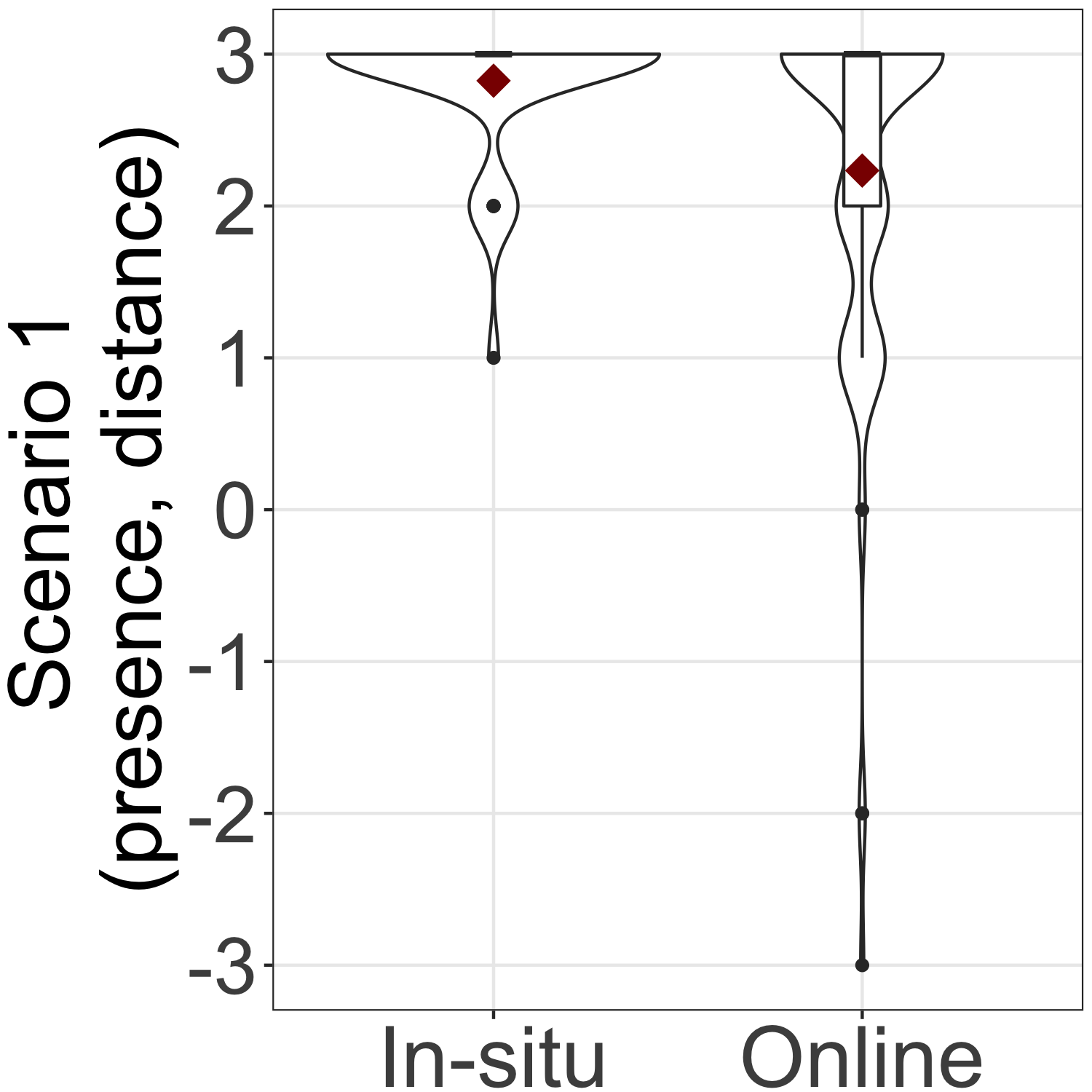}
        \caption{The in-situ sighted were more positive about \textit{Scenario1} than the online sighted.}
        \label{fig:diff_sq1}
    \end{subfigure}%
    \hspace{0.45em}
    \begin{subfigure}[t]{0.155\textwidth}
        \centering
        \includegraphics[width=\textwidth]{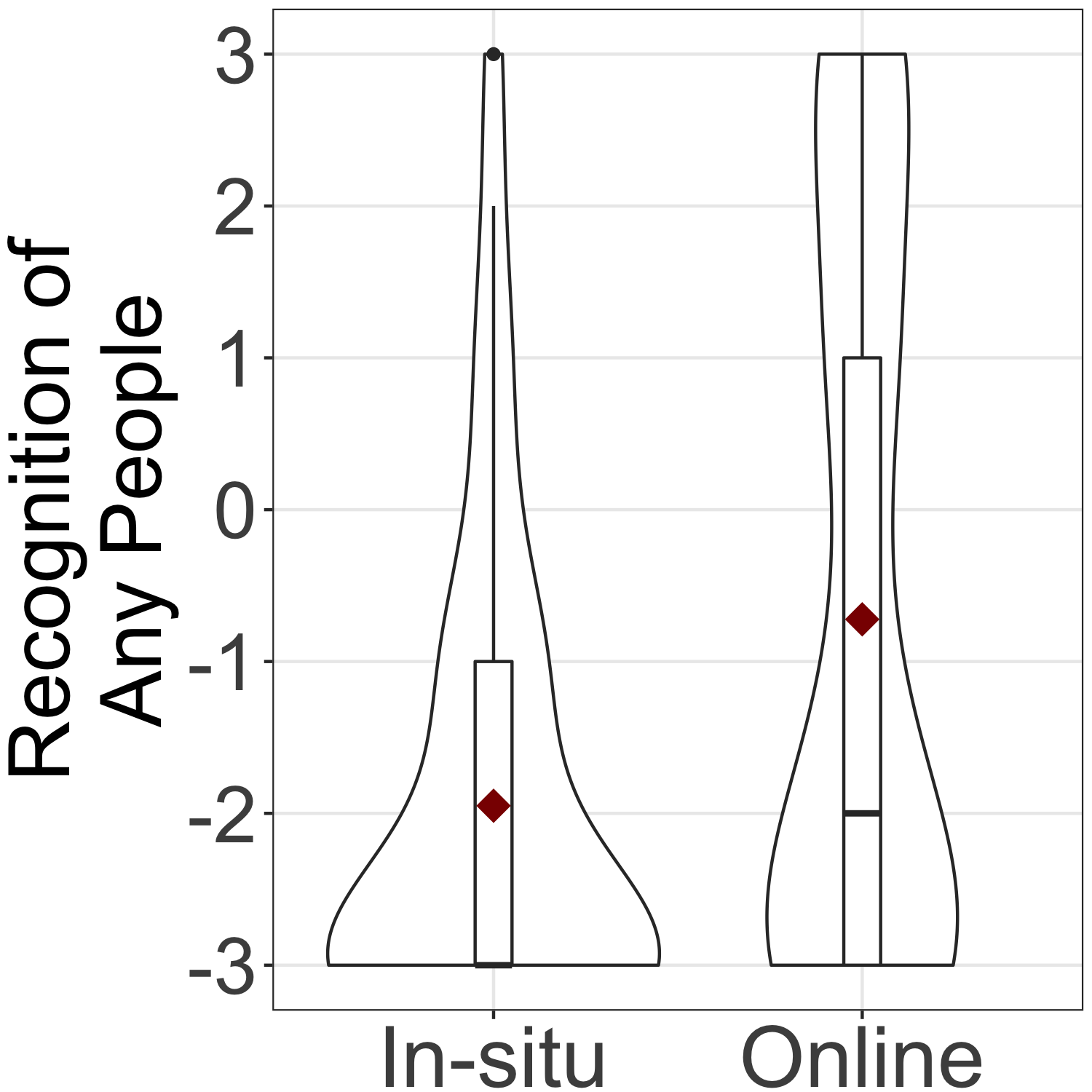}
        \caption{The in-situ sighted were more negative about any people recognition than the online sighted.}
        \label{fig:diff_sq5}
    \end{subfigure}%
    \hspace{0.45em}
    \begin{subfigure}[t]{0.155\textwidth}
        \centering
        \includegraphics[width=\textwidth]{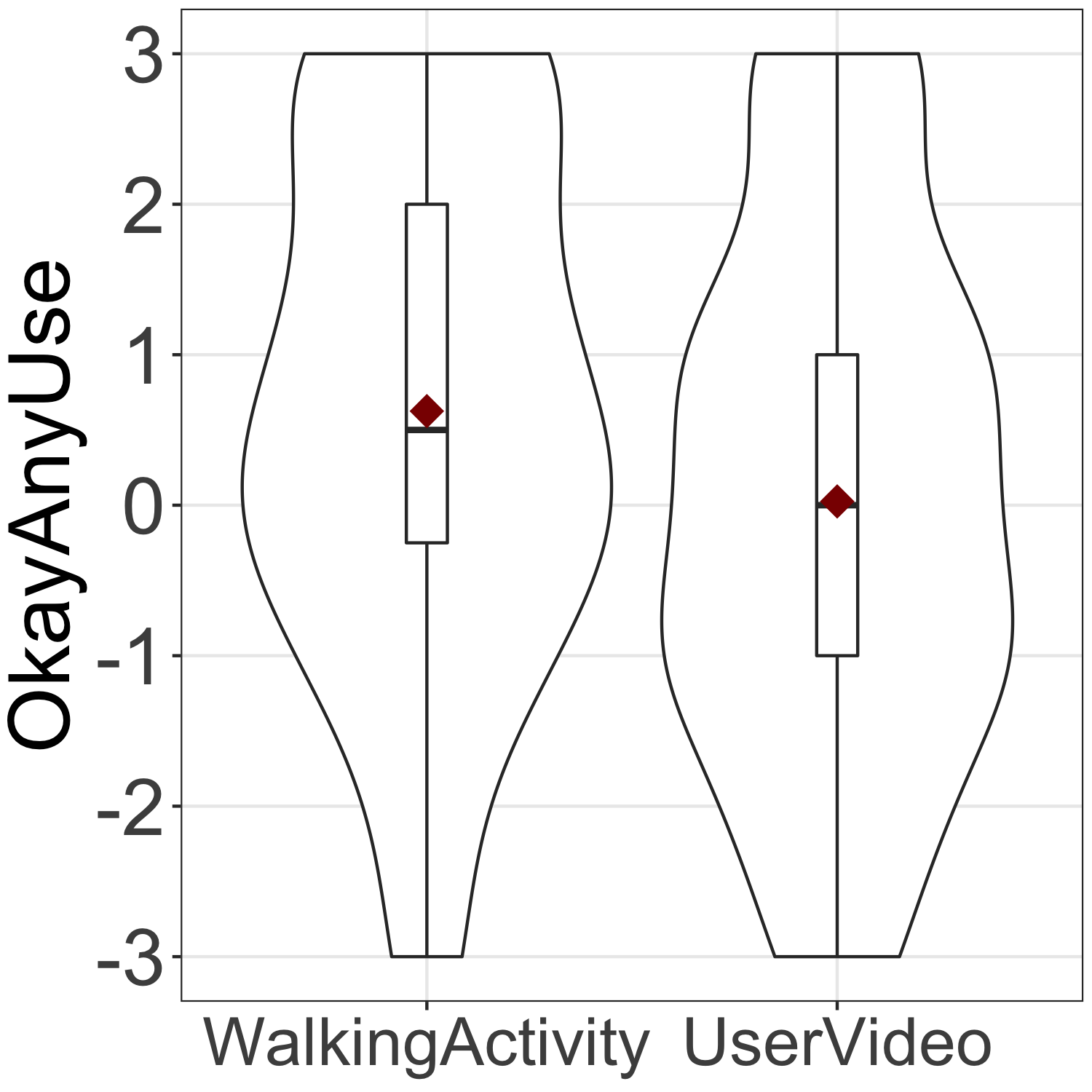}
        \caption{The in-situ sighted became more negative about any people using it after the user video.}
        \label{fig:insitu_any_before_after}
    \end{subfigure}%
    \hspace{0.45em}
    \begin{subfigure}[t]{0.155\textwidth}
        \centering
        \includegraphics[width=\textwidth]{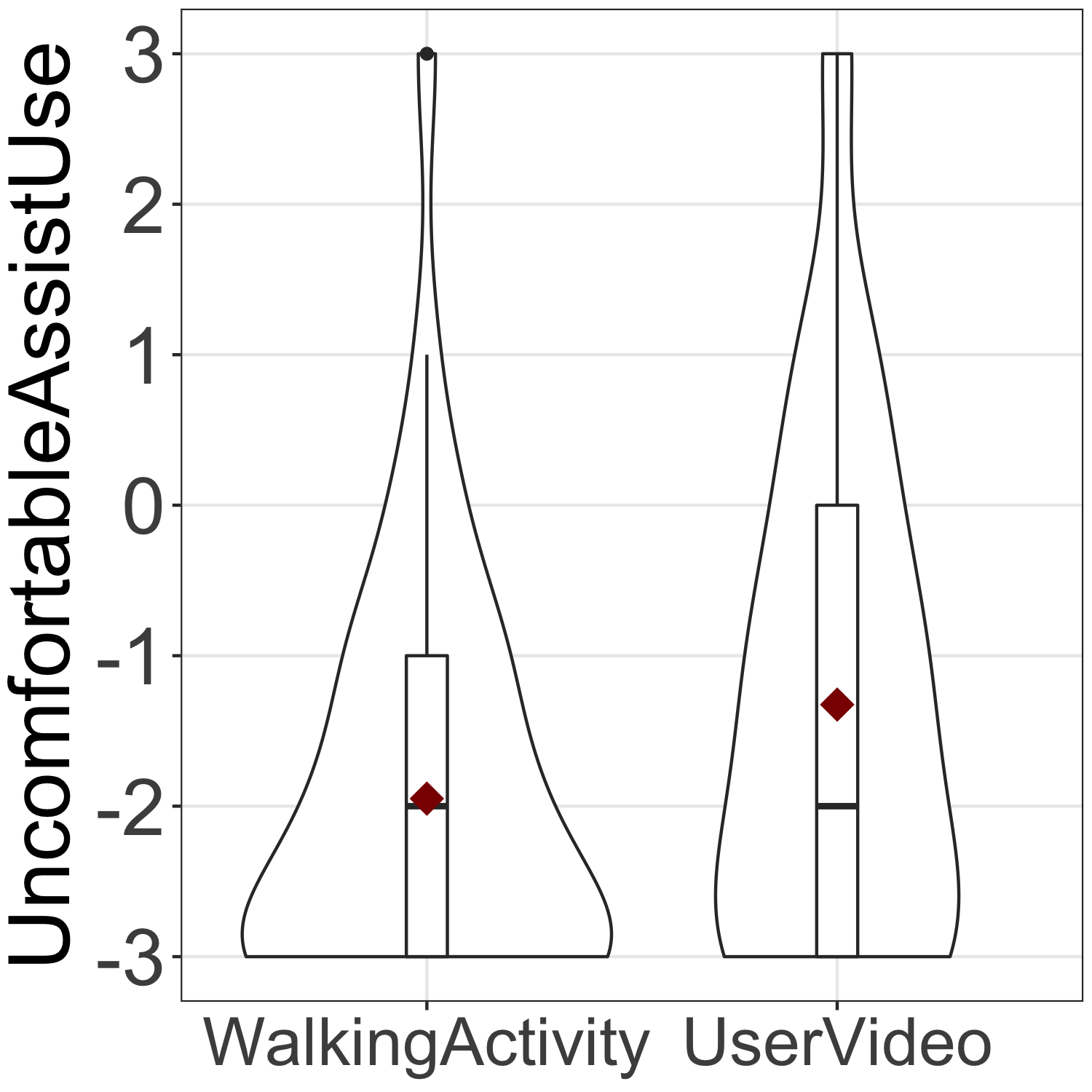}
        \caption{The in-situ sighted felt more uncomfortable about the camera after the user video.}
        \label{fig:insitu_vq4_before_after}
    \end{subfigure}%
    \caption{The four figures on the left for comparison of responses from in-situ sighted participants with those from online sighted participants, and two figures on the right for comparison of in-situ sighted participants' responses before and after the user video.}
    \label{fig:insitu_vs_online}
\end{figure*}

\subsubsection{Perspectives of sighted participants}
We attempted, both in online and in-situ studies, to have sighted people who are interacting often with technology either as Turkers (online) or as students, faculty, and visitors in a technological institution (in-situ), and thus include participants who tend to be positive toward technology in general. Our rationale behind this decision, beyond the obvious practicality, was to reach out to an audience seen as potential early adopters of technology.
Using responses from in-situ and online sighted participants, we further evaluated social acceptance of an assistive wearable camera from perspectives of sighted passerby.
Since the in-situ participants personally experienced the smart glasses in the user study, responses from the online participants who remotely experienced the smart glasses ($n=90$) were used to control variances that can be caused by the camera form.

After the walking activity in the in-situ study, except for \textit{UncomfortableRecording} ($\mu=1.0$, $\sigma=1.5$), the in-situ participants seemed to be generally positive about the assistive technology in the other seven categories: \textit{OkayAssistUse} ($\mu=2.4$, $\sigma=1.0$), \textit{OkayAnyUse} ($\mu=0.6$, $\sigma=1.7$), \textit{PrivacyConcerns} ($\mu=-0.1$, $\sigma=1.8$), \textit{UncomfortableAssistUse} ($\mu=-2.0$, $\sigma=1.4$), \textit{OkayRecording} ($\mu=1.0$, $\sigma=1.8$), \textit{OkaySingleUse} ($\mu=2.6$, $\sigma=0.7$), and \textit{UncomfortableSingleUse} ($\mu=-1.9$, $\sigma=1.6$). 
As for \textit{UncomfortableRecording}, both the in-situ and online (Glasses) participants tended to feel uncomfortable about camera recording (online (Glasses): $\mu=0.3$, $\sigma=2.0$), and there was no significant difference between these two groups.
On the other hand, we found that the in-situ participants had more positive attitudes toward the single use of images/videos in the assistive technology than did the online participants.  As depicted in Figure~\ref{fig:diff_v1q7} and \ref{fig:diff_v1q8}, the in-situ participants were more positive about the assistive technology using images/videos for one-time detection ($p<0.01$, $r=0.272$) and felt less uncomfortable about it ($p<0.05$, $r=0.206$) than the online.


\begin{figure}[t]
    \centering
    \includegraphics[width=0.47\textwidth]{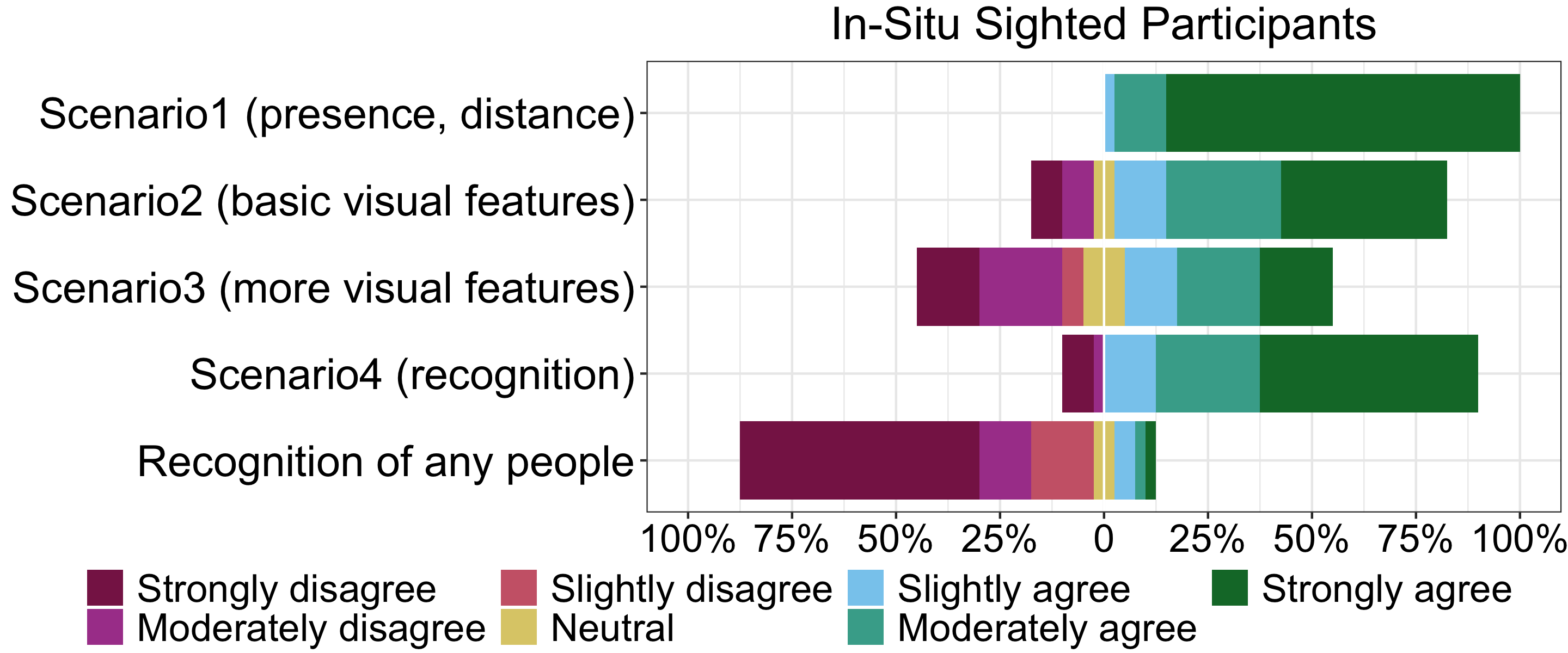}
    \caption{In-situ sighted participants' opinions on use cases of assistive wearable cameras.  Participants were asked whether they would be okay with each of the use cases.}
    \label{fig:insitu_scenarios}
\end{figure}

Moreover, Figure~\ref{fig:insitu_scenarios} visualizes the in-situ participants' responses to the questions about the scenarios introduced in the user video.
We compared these responses with the ones from the online participants (Glasses) to see if there was any significant difference between the groups.
As described in Figure~\ref{fig:diff_sq1}, we found that the in-situ participants were generally more positive toward the technology detecting their presence and distance between blind users and them, \textit{Scenario1 (presence, distance)}, than the online participants were ($p<0.01$, $r=0.273$).
In addition, although both of the groups were negative toward the system recognizing anyone (as opposed to just known people), Figure~\ref{fig:diff_sq5} show that the in-situ participants were more negative about being recognized when they do not know blind users ($p<0.01$, $r=0.246$).

\subsubsection{Awareness of attributes detected by technology}

\begin{figure}[t]
    \centering
    \includegraphics[width=0.45\textwidth]{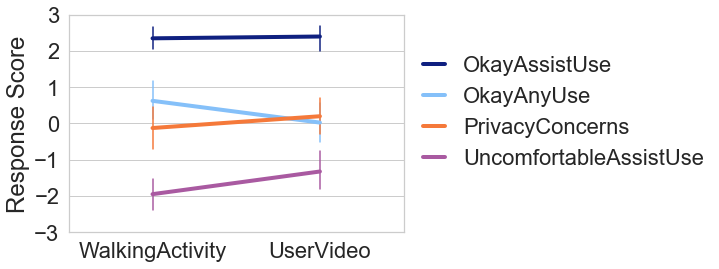}
    \caption{In-situ sighted participants' responses to the social acceptance questions before and after the user video.  Responses are converted to numeric values ($-3$ to $3$): $-3$=\textit{Strongly disagree}, $0$=\textit{Neutral}, $3$=\textit{Strongly agree}.}
    \label{fig:insitu_social_acceptance}
\end{figure}

Analyzing in-situ sighted participants' responses to the social acceptance questions before and after the user video, we also evaluated effects of awareness of the data used in the technology on the in-situ sighted participants' perceptions of the technology; they learned about what such an assistive system may detect from pedestrians by watching the user video in the post-study session.
Figure~\ref{fig:insitu_social_acceptance} shows their responses to the questions before and after the user video.
Similar to our finding from the online survey data, we observed that in-situ sighted participants grew more negative toward anyone using a wearable camera system (\textit{OkayAnyUse}) after the user video ($p<0.05$, $r=0.282$) as shown in Figures~\ref{fig:insitu_any_before_after} and \ref{fig:insitu_social_acceptance}.
Before the user video, one in-situ participant mentioned, ``it doesn't cause any inconvenience to others,'' but, after the user video, that same participant commented, ``they may use it to analyze our personal information for some bad purpose''; this participant changed the response from \textit{Strongly agree} to \textit{Moderately disagree}.

As shown in Figure~\ref{fig:insitu_vq4_before_after}, we found that the in-situ participants felt more uncomfortable about blind people using the assistive camera (\textit{UncomfortableAssistUse}) after watching the user video and realizing that their visual features were detected by the system ($p<0.05$, $r=0.253$), which was not observed from the online participants.
One in-situ participant said ``it's okay for me'' and strongly disagreed on \textit{UncomfortableAssistUse} before watching the user video, but, after the video, mentioned ``when I realize it will collect too much information, I will feel a little bit uncomfortable'' and changed to slightly agreeing to feel uncomfortable about it.

\subsubsection{Blind users and sighted pedestrians}


\begin{figure}[t]
    \centering
    \begin{subfigure}{0.23\textwidth}
        \centering
        \includegraphics[width=\textwidth]{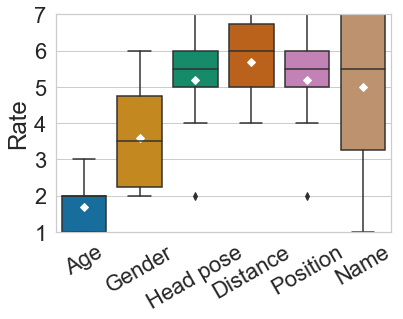}
        \caption{Blind participants' rating for each feature detection.  The top-3 preferred features are distance, head pose, and position.  (1=the least preferred, 7=the most preferred)}
        \label{fig:blind_rate}
    \end{subfigure}%
    \hspace{0.5em}
    \begin{subfigure}{0.23\textwidth}
        \centering
        \includegraphics[width=\textwidth]{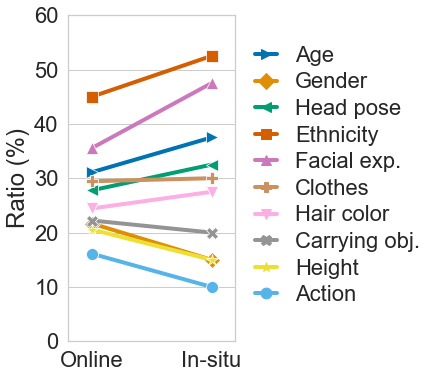}
        \caption{Ratio of sighted participants for each feature that do not like to be detected by such assistive system.}
        \label{fig:sighted_ratio}
    \end{subfigure}%
    \caption{Participants' opinions on feature detection.}
    \label{fig:feature_pref}
\end{figure}

Lastly, we further investigated conflicts of interest in functionalities of an assistive wearable camera from the perspectives of blind users and sighted passerby.
Blind participants were asked to rate six visual features (age, gender, head pose, distance, position, name) that the prototype system provided them with during the user study.  The rating scale was 1 to 7 where 1 indicated the least important feature and 7 indicated the most important feature; redundancy was allowed for the rating.
On the other hand, we asked sighted participants which of their visual features they would prefer not to be detected and reported by such an assistive system.

Figure~\ref{fig:blind_rate} shows that blind participants preferred to obtain information about \textit{head pose}, \textit{distance}, and \textit{position} of pedestrians, while \textit{age} is the last feature that blind participants did care about.
P4 mentioned, ``if there is a baby one year old, (the baby) can't tell you what the number on a door is.  Other than that, (age estimation) doesn't make any difference.'' 
Figure~\ref{fig:sighted_ratio} shows the ratio of sighted participants for each feature that they would not want be exposed by the assistive system. \textit{Ethnicity}, \textit{facial expression}, and \textit{age} are the top-3 features that participants would not want such a system to detect from them.
Yet, some sighted participants (31\% of online sighted participants and 23\% of in-situ sighted participants) answered that they would allow an assistive system to detect any necessary visual features from themselves.

Features that blind participants did not care to obtain from such an assistive system mostly match those that sighted participants did not want to provide via the system; \textit{age} and \textit{ethnicity} are the features that sighted participants did not want the system to detect, and blind participants did not think that the system needed to provide.
On the other hand, most of the blind participants considered \textit{head pose} as an essential feature that such a system needed to provide because they can use this information as an indicator for pedestrian's availability for social interactions.  In particular, P10 remarked, ``now people with cell phones, they are talking on their phones, but (blind people) don't know whether they are talking to you or on a phone.  The key is if they are looking at (blind people).'' However, around 30\% of each group of sighted participants did not want to provide this information to the system.
Also, it seems that \textit{facial expression} could be an additional indicator for blind users to decide whether to begin social interactions with pedestrians; P5 said, ``it might be nice to have facial expressions.  If someone is upset, I might not want to ask them for assistance.''
However, \textit{facial expressions} was the second feature that sighted participants (36\% of the in-situ and 48\% of the online) did not want the system to detect from them.

\section{Discussion}



We disclose that there were conflicting features that blind users would need from the assistive technology but sighted pedestrians might not want be detected by assistive wearable cameras. These include head pose and facial expression.
In addition, some of the blind participants commented that they do not want such a system to be an indication of their disability; P7 mentioned, ``I wouldn't wear heavier, bulkier smart glasses because it's more likely to draw unwanted attention.''  On the other hand, our analysis with sighted participants indicates that camera visibility and awareness of the technology may affect pedestrians' perceptions of such an assistive wearable camera; the less visible a camera was, the more negative the response of sighted participants toward the technology became after discovering its image/video use.
One sighted participant said, ''I'm not looking to be recorded without my knowledge, so it's important that it is discernible that the blind individual is wearing a camera.''.
It seems as if the more visible camera might have caused people to prepare for being recorded. The less visible camera (smart glasses), on the other hand, might have led sighted people to unpleasantly learn about its photo-taking or video recording function after they already had been recorded.
Hence, conflicts of interest in such technology should first be collected, carefully reviewed, and appropriately addressed for all stakeholders.

Our analysis on responses from online and in-situ sighted participants provides further evidence on the prior findings that people are willing to accept wearable camera technology for assistive purposes~\cite{profita2016effect}.  However, the differences in the online and in-situ sighted participants' responses to \textit{Recognition of any people} indicate that in-person experience may engender different effects on people's perception of such technology than remote experiences do.  It indeed reconfirms the prior finding that in-situ (direct) experience catalyzes people's perception of technology more than remote (indirect) experience did in the user study~\cite{duerden2010impact}. Although we acknowledge the high cost of conducting on-site user study, we encourage researchers to combine on-site user studies with online user studies to better understand people's opinion on technology.

There are, however, several limitations to our work.
First, our in-situ participants might not represent the true population due to their skewed demographic information, such as age and educational background. Among both blind and sighted participants in the in-situ user study, the age distribution was skewed (blind participants: 49--73 years, and sighted participants: 19--39 years).
We see similar trends in their attitudes toward assistive wearable camera technology even when controlling for educational background and comparing the 40 in-situ sighted participants to 53 online sighted participants, who watched the smart glasses videos and indicated having a bachelor's or advanced degree, but it may not indicate that there is no demographic or confounding factor that may have led to the differences in social acceptance between the two groups of sighted participants.  We believe that conducting a user study with more participants with various backgrounds might address such issues and help us understand social acceptance, more generally.

Moreover, the scenario of detecting pedestrians in a corridor was chosen to be the main use case for our working prototype system, since it allowed us to conduct user study and measure social acceptance of such technology in a controlled environment. This single scenario enables repeatability of our user study, but could limit exploration of other factors in social acceptance due to its lack in representability of a real-world use case. As such technology could reveal different social tensions in different scenarios, our future work will be to explore further use cases of assistive wearable cameras to evaluate its social acceptance in general.





\section{Conclusion}
Rise of wearable camera technology has triggered researchers' attention to its social acceptance and potential assistive uses.
Considering not only users and bystanders but also several other factors that may affect people's perceptions of such technology, our work explored social acceptance of a potential use, \textit{pedestrian detection}, of a wearable camera system for blind people.
We found that there was agreement and disagreement between blind users and sighted pedestrians on what such a system should and should not provide, and observed that both the camera visibility and the way of experiencing the technology may affect bystanders' attitudes toward the technology. We also reconfirmed that people were more positive toward wearable cameras if it was designed for assistive purposes.
Our findings suggest that it would be necessary to evaluate social acceptance of such technology in a more realistic environment and understand the needs of not only users but also people who would be involved in the technology.

\section{Acknowledgments}
We thank all participants who took part in our user studies as well as our anonymous reviewers for their insightful feedback on an earlier version of this paper.
This work is supported by Shimizu Corporation and NIDILRR (\#90REGE0008).

\balance{}

\bibliographystyle{SIGCHI-Reference-Format}
\bibliography{proceedings}

\end{document}